\documentclass[%
groupedaddress,
aps,
pre,
rsi,%
amsmath,amssymb,
reprint,%
]{revtex4-1}
\usepackage{graphicx}
\usepackage{dcolumn}
\usepackage{bm}
\usepackage[export]{adjustbox}
\begin{document}


\title{Ultrametricity of optimal transport substates for multiple interacting paths over a square lattice network}

\author{Marco Cogoni}\email{marco.cogoni@crs4.it}
\author{Giovanni Busonera}\email{giovanni.busonera@crs4.it}
\author{Gianluigi Zanetti}\email{gianluigi.zanetti@crs4.it}
\affiliation{CRS4: Center for Advanced Studies, Research and Development in Sardinia - 09010 Pula (CA), Italy}

\date{\today}

\begin{abstract}
We model a set of point-to-point transports on a network as a system of polydisperse interacting self-avoiding walks (SAWs) over a finite square lattice. 
The ends of each SAW may be located both at random, uniformly distributed, positions or with one end fixed at a lattice corner.
The total energy of the system is computed as the sum over all SAWs, which may represent either the time needed to complete the transport over the network, or the resources needed to build the networking infrastructure.
We focus especially on the second aspect by assigning a concave cost function to each site to encourage path overlap. A Simulated Annealing optimization, based on a modified BFACF algorithm developed for polymers, is used to probe the complex conformational substates structure at zero temperature.
We characterize the average cost gains (and path-length variation) for increasing polymer density with respect to a Dijkstra routing and find a non-monotonic behavior as recently found for random networks.
We observe the emergence of ergodicity breaking and of non-trivial overlap distributions among replicas when switching from a convex to a concave cost function (e.g., $x^\gamma$, where $x$ represents the node overlap).
Finally we show that the space of ground states for $\gamma<1$ is compatible with an ultrametric structure as seen in many complex systems such as some spin glasses.
\end{abstract}

\pacs{Valid PACS appear here}


\maketitle

The problem of optimal transport over various kinds of networks is important both for theoretical and practical reasons~\cite{JayanthR1999}. Areas of application range from river networks~\cite{Rinaldo2014,Ijjasz-Vasquez1993} to vascular systems in animals and plants \cite{Katifori2010,Corson2010}, and from electric energy distribution systems \cite{Bohn2007} to communication networks~\cite{Banavar2000,Yeung2013a}. The adoption of a cost function minimization scheme has allowed a unified approach to very diverse research fields. Cost functions may be thought as energy dissipation for electricity grids, time delay and/or resources needed to build the networking infrastructure. In recent years, the relation between the properties of the cost function and the associated optimal solutions has been extensively studied~\cite{Banavar2000}. For instance, when multiple sources are connected to a single destination (as in drainage basins), it is known that a concave cost leads to multiple (nearly equivalent) spanning trees, whereas a convex behavior shows a unique redundant solution with many loops. When the character of the cost function is not well defined, no \textit{a priori} conclusions may be drawn~\cite{Yeung2012}.

It is well known \cite{Banavar2000} that, for concave cost functions, a multiplicity of local optimal solutions exists. 
Since a hierarchical organization of the states is observed, one may hypothesize an ultrametric relation among them. Ultrametricity (UM) is one of the key features of the mean-field Parisi picture for spin glasses\cite{mezard1987spin}: the states of the system obey an UM distance which translates into a hierarchical organization.
This behavior has been hypothesized or observed for different polymer systems with noise, such as directed polymers in random media (DPRM)~\cite{Zhang1987, Kardar-Zhang1987,Pang-Healey1993} and for self-interacting SAWs in external fields \cite{fernandez1991ultrametricity}.

In this paper we propose a model having three main differences with respect to the DPRM formulation: 1) the (locally) minimum cost is achieved by collectively optimizing several interacting chains; 2) polymers are polydisperse and have at least one end which is randomly located on the lattice meaning that quenched disorder is achieved through random topology rather than noisy bonds; 3) our polymers may not be directed: they are free to wander backwards/sideways to achieve a global cost gain. This behavior is seldom observed for weakly interacting polymers or very dilute systems.

Our analysis reveals a non-monotonic behavior of the optimized cost gain, with respect to a Dijkstra routing, when increasing polymer density as previously found for random networks~\cite{Yeung2013a}. Moreover, we observe the emergence, only for concave costs, of ergodicity breaking and of non-trivial overlap distributions among replicas. Finally, we present evidence that the space of ground states is compatible with an ultrametric structure.

\paragraph*{Model and Numerical Methods --}
We consider a square lattice network of $N=L^2$ nodes and side $L$, each node connected to its four nearest neighbors via uniform links with adjacency matrix $A_{ij} = A_{ji} = 1$, zero elsewhere. A set of $M$ communications, modeled as polymers with fixed ends, compete on the network for the available resources, each occupying a path described by an interacting SAW. The self-avoidance condition enforces that no path uses the same node more than once, whereas distinct polymers can use the same node. The occupation number of each path $\nu$ on the node $i$ is denoted by $\sigma_i^\nu \in \{0,1\}$. The total occupation number on the $i$-th node is $I_i=\sum_\nu \sigma_i^\nu$. The interaction among polymers is regulated by a Hamiltonian of the form \begin{equation}
H=M\sum_i f(k_i {I_i\over M})
\end{equation}
where the concave/convex character of the cost function $f$ makes the
system behave in qualitatively different ways~\cite{Banavar2000}. The occupation number is normalized by $M$ in order to have a uniform temperature behavior with respect to polymer multiplicity.
The $k_i$ are node-dependent weights, in general set to unity, which may be used to model spatial non-uniformity. In this paper we consider only $k_i=1$ and the simplest functional form for the cost $f$: the power function $f(x)=x^\gamma$. This functional dependence leads to polymer repulsion for $\gamma>1$, whereas $\gamma<1$ encourages overlap. For $\gamma=1$ the polymers do not interact, and it is known that the ground state of $H$ is attained by (highly degenerate) shortest-path routing~\cite{Yeung2013a,Banavar2000}. 

To explore the energy landscape of $H$ for $\gamma\neq1$, we adopt a Simulated Annealing (SA) scheme in which temperature is gradually decreased to zero within a canonical Monte Carlo (MC). The basic MC move follows the BFACF algorithm developed for lattice polymers~\cite{BFACF,BFACF2}. With respect to the original scheme, at each iteration one polymer is randomly selected and instead of applying the basic move at a single random site, we perform multiple moves. The number of moves on the SAW is randomly chosen within one and the average polymer length. This random choice of basic move multiplicity guarantees good MC acceptance rates~\cite{Sergio1990}. Since in this work we are dealing with several interacting polymers, we rely on the Gibbs acceptation factor of the MC to extract a chain of states from the canonical ensemble instead of direct generation as in Ref. \cite{BFACF}. Assigning the same probability both to path-enlarging and path-shrinking BFACF moves leads to very low acceptance rates. We solve this problem by tying the probability of path-enlarging moves to the MC temperature~\cite{binder1995monte}.

The square lattice with uniform $k_{i}=1$ factors induces on this problem some peculiarities not found in a continuous representation of space. In particular, the ground states for basic routing problems (i.e., involving non-interacting polymers) in two dimensions are intrinsically degenerate because ${{m+n}\choose{n}}$ solutions with the same energy exist between any pair of points with $m$ horizontal and $n$ vertical distance \cite{Mernik_anefficient}. We restrict our study to two-dimensional lattices with no periodic boundary conditions (PBC) to better model realistic network topologies. Preliminary numerical results show that qualitatively similar conclusions may be drawn when PBC are imposed at the boundaries.

Detecting UM in finite-volume systems can be very difficult due to finite-size effects especially with no PBC~\cite{Rammal1986,Hed2004,Katzgraber2009,Katzgraber2012, Ciliberti2004}.
To measure differences between replicas $\alpha, \beta$ for the same quenched disorder, we define a path overlap $q_{\alpha\beta}$, computed as the ratio of the common visited nodes (node-overlap $\dot{q}_{\alpha\beta}$) or common visited links (link-overlap $\tilde{q}_{\alpha\beta}$) with respect to path length~\footnote{Whenever two strings possess different lengths, we concatenate a padding string to the short one.}.
For homologue instances of the SAW $\nu$ belonging to replicas $\alpha$ and $\beta$, we define the node overlap as
\begin{equation}
\dot{q}_{\alpha\beta}^{\nu} = {1\over N_{nodes}} \sum_i \sigma_i^{\alpha,\nu} \sigma_i^{\beta,\nu}
\end{equation} 
and the link overlap
\begin{equation}
\tilde{q}_{\alpha\beta}^{\nu} = {1\over N_{links}} \sum_{i,j} A_{ij}\sigma_i^{\alpha,\nu} \sigma_j^{\alpha,\nu} \sigma_i^{\beta,\nu} \sigma_j^{\beta,\nu}.
\end{equation} 
We obtain compatible results for $\tilde{q}$ and $\dot{q}$, but the data shown in this paper is computed by using the link-overlap, so we define $q_{\alpha\beta}\equiv\tilde q_{\alpha\beta}$. 
From the overlap we get the normalized Hamming distance as $\delta_{\alpha\beta}=1-q_{\alpha\beta}$. 

\begin{figure}[htb]
	\centering
	\includegraphics[width=0.45\textwidth]{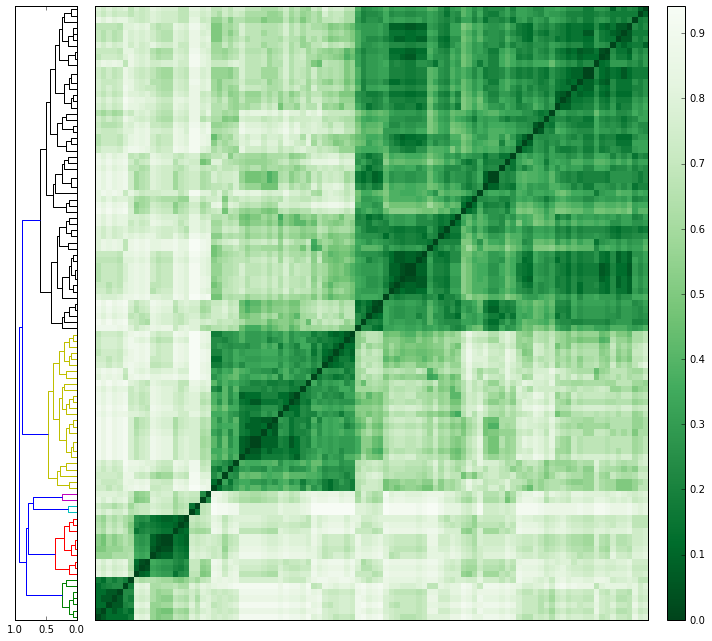}
	\includegraphics[width=0.44\textwidth]{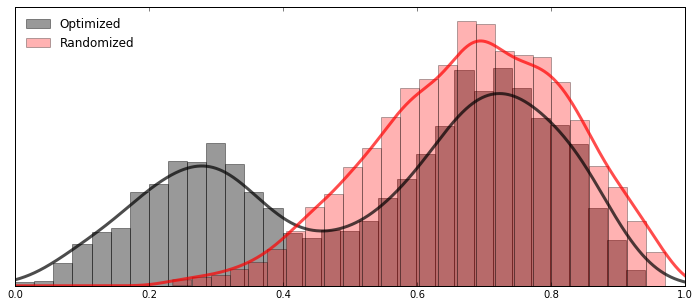}
	\caption{Top: Dendrogram and distance matrix for $100$ states with $N=1024, M=64$ and $\gamma=0.5$. Each matrix element represents the Hamming distance, normalized by path length, for one of the $M$ homologue SAWs belonging to each pair of states. Bottom: The non-trivial distribution of distances in the above matrix (black, optimized polymers); red, pseudo-normal distribution for randomized polymers.
	We obtained similar results for a large fraction of SAWs except for pathological cases. Refer to the Supplemental Materials for details.}
	\label{fig:dendro_matrix1}
\end{figure}

In an UM space \cite{Rammal1986} the triangle inequality $\delta_{\alpha\gamma} \le \delta_{\alpha\beta}+\delta_{\beta\gamma}$ valid for metric spaces, is replaced by a stronger version $\delta_{\alpha\gamma} \le max\{\delta_{\alpha\beta}, \delta_{\beta\gamma}\}$. This inequality is equivalent to imposing that any triple of points should form an acute isosceles triangle or, at most, an equilateral one.
In order to discern between trivial UM due to equilaterals or true UM due to acute isosceles, after performing a standard UM test~\cite{Hed2004}, we propose a new procedure that consists of analyzing the frequency distribution of each triple of ordered distances $\delta_{max}, \delta_{med}, \delta_{min}$, by keeping only two transformed components as defined in Ref.\cite{Contucci2007}: $Y=\delta_{max}-\delta_{med}$ vs $X=\delta_{med}-\delta_{min}$.

We compute the distance $\delta_{\alpha\beta}$ for all pairs of states within the same quenched disorder. This is achieved by selecting pairs of homologue SAWs belonging to two states $\alpha$ and $\beta$, and finally computing their normalized Hamming distance.
By applying a clustering algorithm~\cite{Hed2004,Katzgraber2009,Ciliberti2004} to each disorder, we obtain a dendrogram such as the one in the left part of Fig.~\ref{fig:dendro_matrix1}. The procedure starts with each state in a separate cluster, then iteratively the nearest clusters are merged. At each step the inter-cluster distances are recomputed by averaging among all pairs of their member states. The procedure ends when a root common cluster appears.
We then sort all states in the same order as the dendrogram and finally plot the adjacency matrix: a well-visible block-diagonal complex structure emerges for $\gamma=0.5$ (we found similar results for $\gamma=0.99$), along with a deep hierarchy visible in the dendrogram. On the other hand, nearly flat hierarchies, somewhat uniform distance matrices and pseudo-normal overlap distributions are observed for $\gamma\ge1$ (see Fig. S7).
To probe for an UM space structure, we randomly select three configurations from the hierarchical cluster structure (see Ref.~\cite{Hed2004}), resulting in three mutual distances that we sort to get $\delta_{max}\geq \delta_{med}\geq \delta_{min}$ and finally compute the correlator $K={{\delta_{max}- \delta_{med}}\over \delta_{min}}$. If the phase space is UM, we expect $\delta_{max} \simeq \delta_{med}$ for $L\rightarrow\infty$. Thus $P(K)$ should converge to a delta function in $K=0$ for $L\rightarrow\infty$ and the variance of the distribution $Var(K)\rightarrow 0$. By this approach, UM would be detected even in the case of equilateral triangles (trivial UM). The alternative definition of $K$ in Ref. \cite{Katzgraber2009} is difficult to apply for polydisperse SAWs (each SAW length has a distinct distance distribution), so we devised a supplementary test to rule out trivial UM.

\paragraph*{Results and discussion  --}

All ground states have been obtained by performing a SA energy minimization with an exponential cooling converging to $T=0$ at one third of the total simulation length. To characterize cost gain and path-length variation for the system with both random ends, we performed several minimizations in which the number of SA timesteps varied with system size from $0.5\cdot10^6$ for $L=8$ and $M=8$ to $10\cdot10^6$ for $L=64$ and $M=1024$. The maximum number of basic BFACF moves per timestep was proportional to $\sqrt{N}$. For each lattice size we generated $10$ quenched disorders with random uniform configurations and for each disorder we produced $10^2$ local ground states. We plan to expand this number for future works. Some examples of the ground states we obtained are shown in Figs. S1-S4.

\begin{figure}[htb!]
	\includegraphics[width=0.9\linewidth]{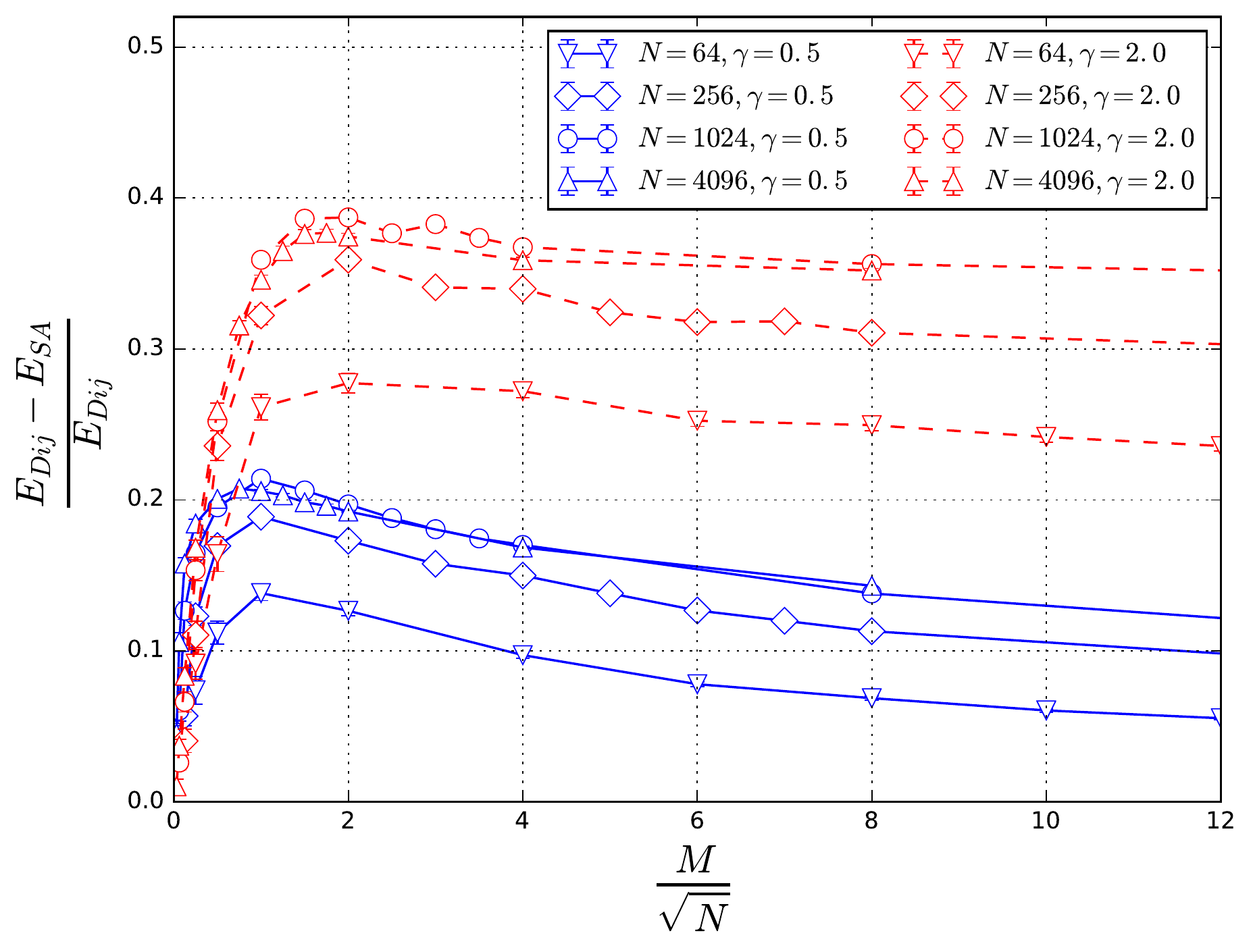}
	\caption{Cost decrease ratio of the SA with respect to a Dijkstra algorithm.}
	\label{fig:scaling_costs}
\end{figure}

The first goal has been to characterize the system as regards the attainable cost gain with respect to the shortest path routing that is widely used for many transport applications. In Fig.~\ref{fig:scaling_costs} we plot the energy difference ratio both for concave ($\gamma<1$, continuous blue online) and convex ($\gamma>1$, dashed red online) cost functions.

\begin{figure}[htb!]
	\includegraphics[width=0.93\linewidth]{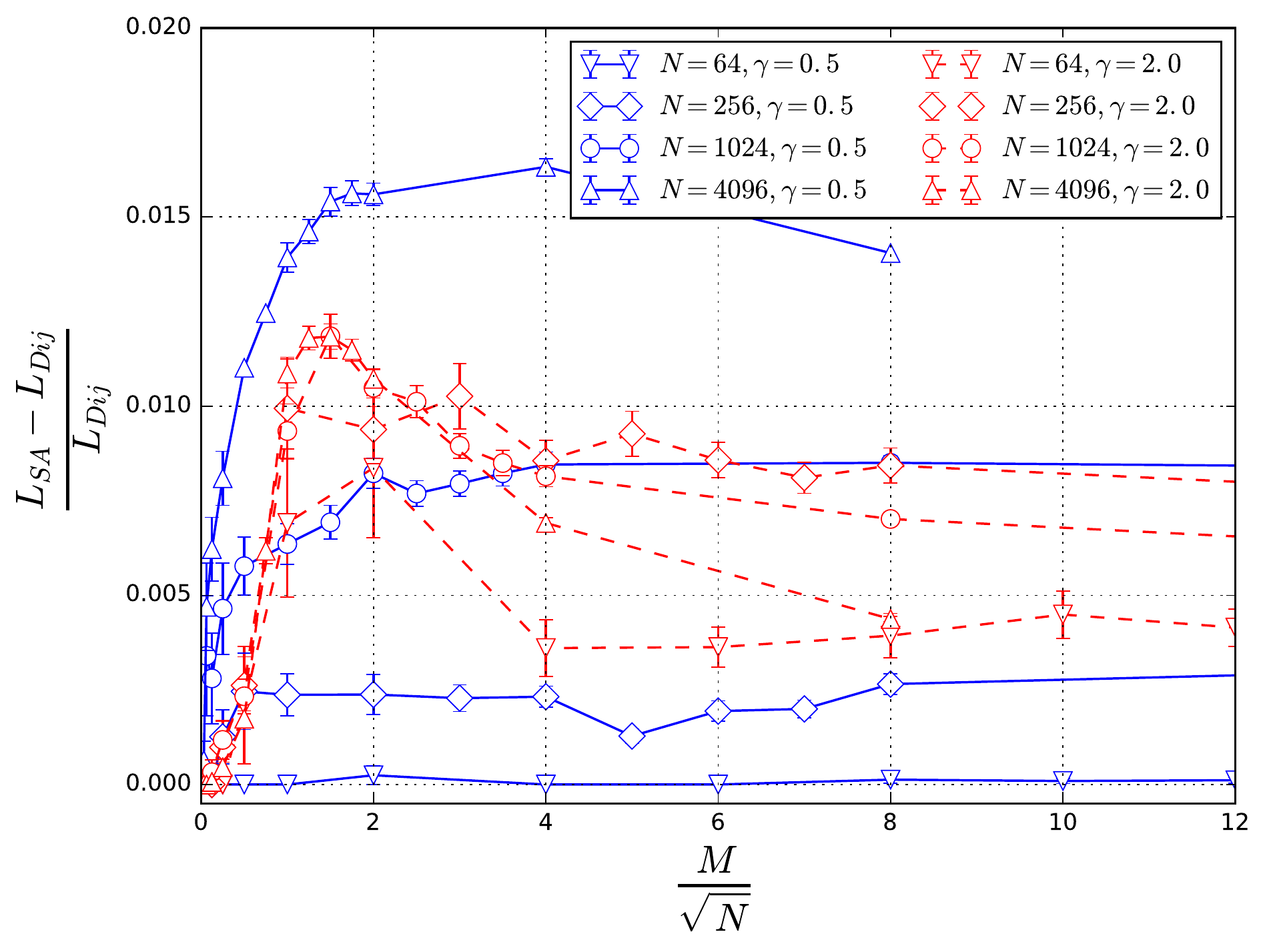}
	\caption{Path length increase ratio for the SA with respect to a Dijkstra algorithm.}
	\label{fig:scaling_costs_path}
\end{figure}

Since the average polymer length increases with $L$, we observe in Fig.~\ref{fig:scaling_costs} the tendency of the curves to superimpose for $N\rightarrow\infty$. After a steep cost gain growth, maximum efficiency with respect to Dijkstra is reached for both $\gamma$ values, then the value slowly decreases since most nodes are already busy and the advantage of longer-than-Dijkstra detours is weaker. The peaks are shifted for $\gamma=0.5$ and $\gamma=2.0$: $M/\sqrt{N}\sim1$ and $\sim2$, respectively. Cost gain ratios are quite different in the two scenarios: for the convex case, the gain ratio is relatively constant for any $M$ at nearly $40\%$, while in the concave situation, it rarely goes beyond $20\%$ and decreases more markedly for higher $M$ values.
By comparing the peak values of Fig.~\ref{fig:scaling_costs} with their associated path-length variations in Fig.~\ref{fig:scaling_costs_path}, we show that large cost gains may be obtained by employing paths slightly longer than Dijkstra ($< 2\%$ for both $\gamma$ values).
These results are qualitatively very similar to those obtained by an alternative optimization approach presented in Refs.~\citep{Yeung2012,Yeung2013a} for random paths on a random graph with constant connectivity $k=3$. It should be stressed that the regular lattice is not tractable with that method since too many loops exist, leading to severe ground state degeneration. The present method may be exploited to minimize network construction resources by obtaining a set of lean structures with $\gamma<1$ (resource sharing encouraged) and then selecting the best performing ground states for $\gamma>1$ (paths competing for resources).

Let us focus for the rest of the paper on the concave cost case. In the Supplementary Materials to Ref.~\cite{Yeung2013a}, there is a brief discussion regarding the possibility of a Replica Symmetry Breaking (RSB) scenario for $\gamma<1$. This led us to question whether an UM structure among the ground states of our closely related system exists. We consider two polymer distributions: one in which both SAW ends are uniformly distributed and another in which one end is constrained to a lattice corner. The RSB scenario is apparent when looking at non-trivial overlaps~\cite{Franz-Parisi2000} among replicas at $T=0$: we observe multimodal overlap distributions for roughly half of the SAWs. This fraction grows with $N$. In the SM we compare individual dendrograms, distance matrices and distributions. For randomized paths and for $\gamma\ge1$ the overlap distributions are always quasi-normal.

\begin{figure}[htb]
\includegraphics[width=\linewidth]{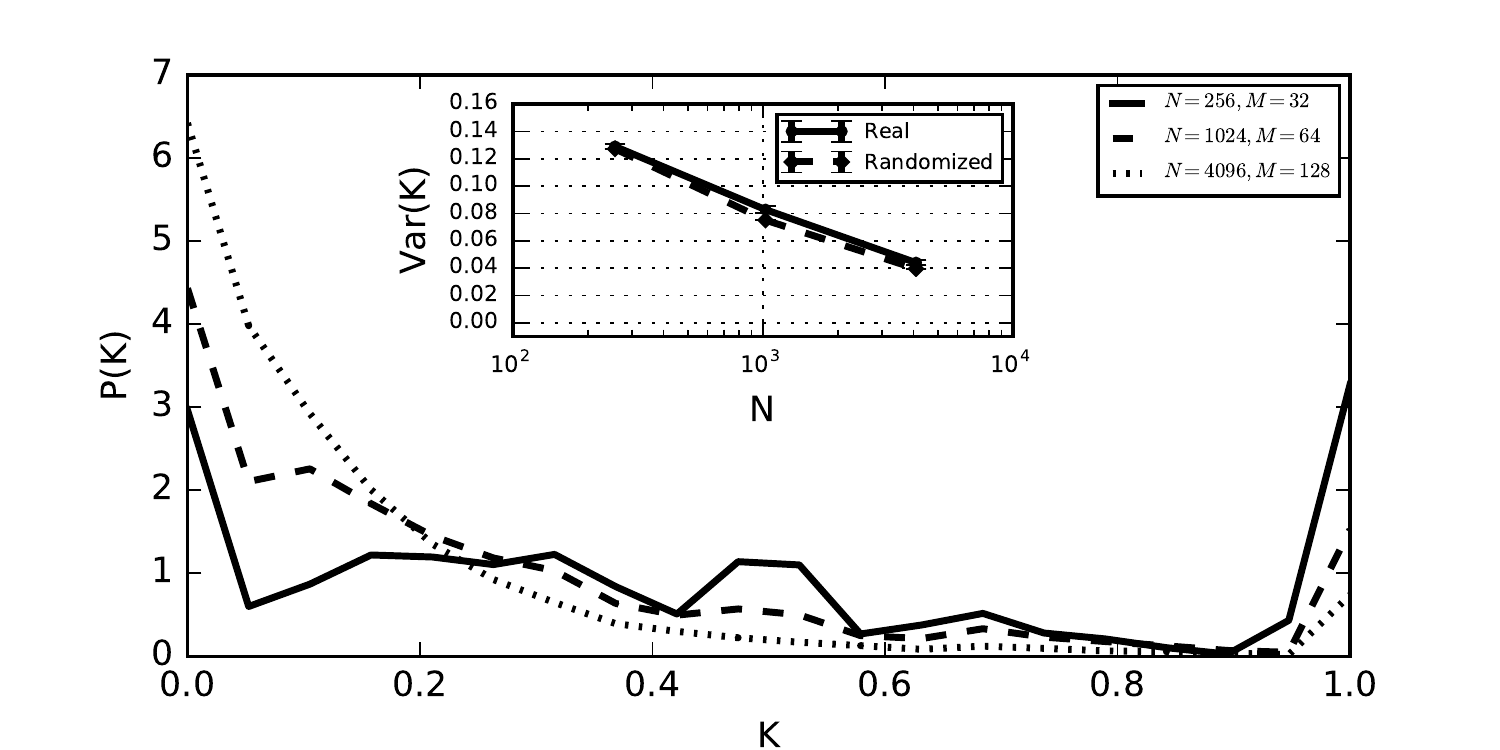}
\includegraphics[width=\linewidth]{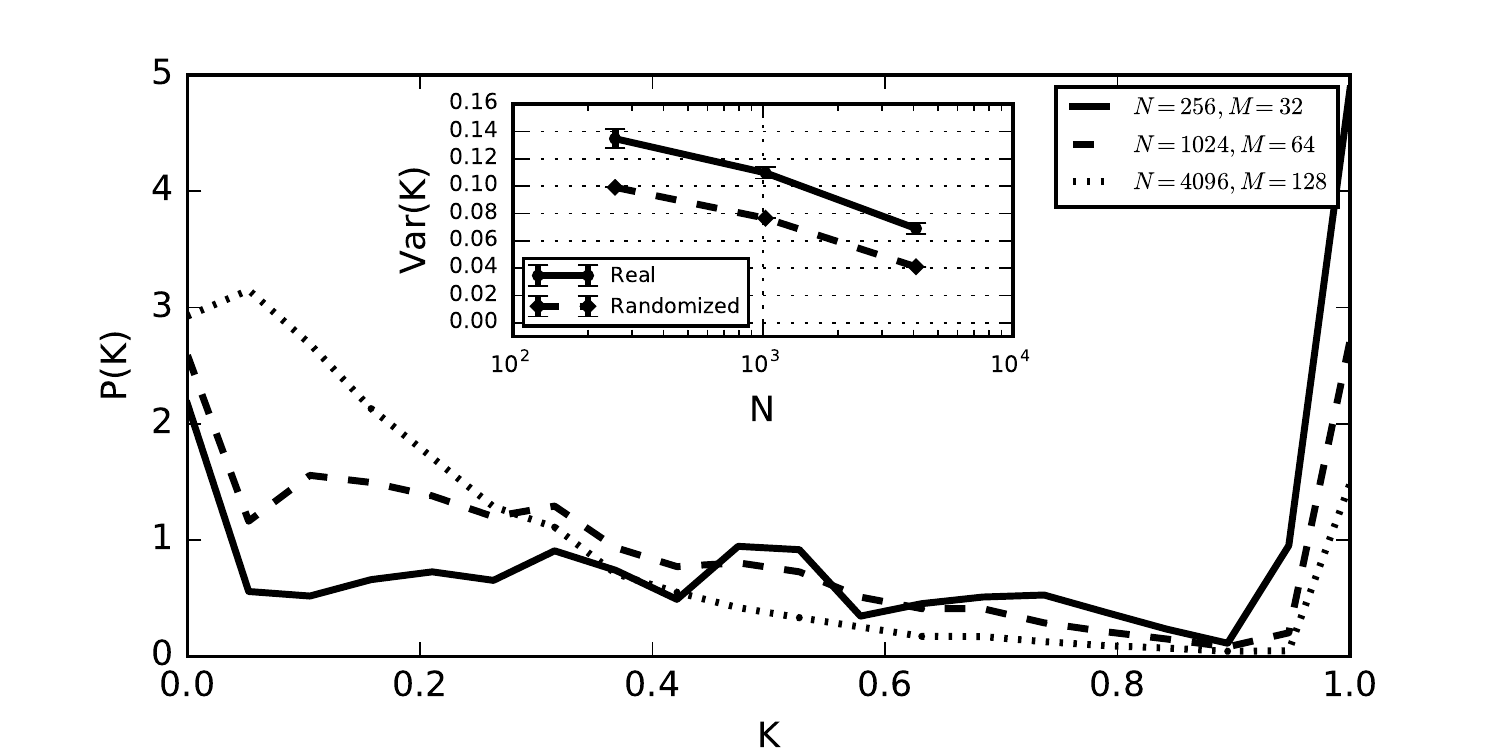}
\caption{Distribution $P(K)$ for different $N$ and polymer number $M=2\sqrt{N}$. Top graph refers to the unconstrained polymers, while bottom to the corner-constrained case. The peak associated to collinear triangles ($K\approx 1$) tends to disappear with growing $N$ for both systems.
}
\label{fig:pdk_vark}
\end{figure}

It has become customary to show a tendency towards UM by plotting the distribution $P(K)$ for several system sizes along with its variance $Var(K)$. In Fig.~\ref{fig:pdk_vark} the $P(K)$ for the random-ends polymers (top) and for the fixed-end polymers (bottom) are shown. There is a visible trend of both $P(K)$ to diverge at $K\rightarrow 0$ when $N\rightarrow\infty$. Both polymer topologies share the same overall behavior.
In the two insets of Fig.~\ref{fig:pdk_vark}(top and bottom) we can compare the variance of the $P(K)$ for both systems compared to their randomized counterparts~\footnote{The randomized system is obtained by reshuffling every SAW sequence with the same quenched disorder. The randomized paths do maintain legal connection between polymer ends, but self-avoidance may be violated.}: they both tend to zero as lattice size grows, so we cannot still conclude in which case whether UM is attained in the thermodynamic limit.

\begin{figure}[htb]
\includegraphics[width=0.99\linewidth]{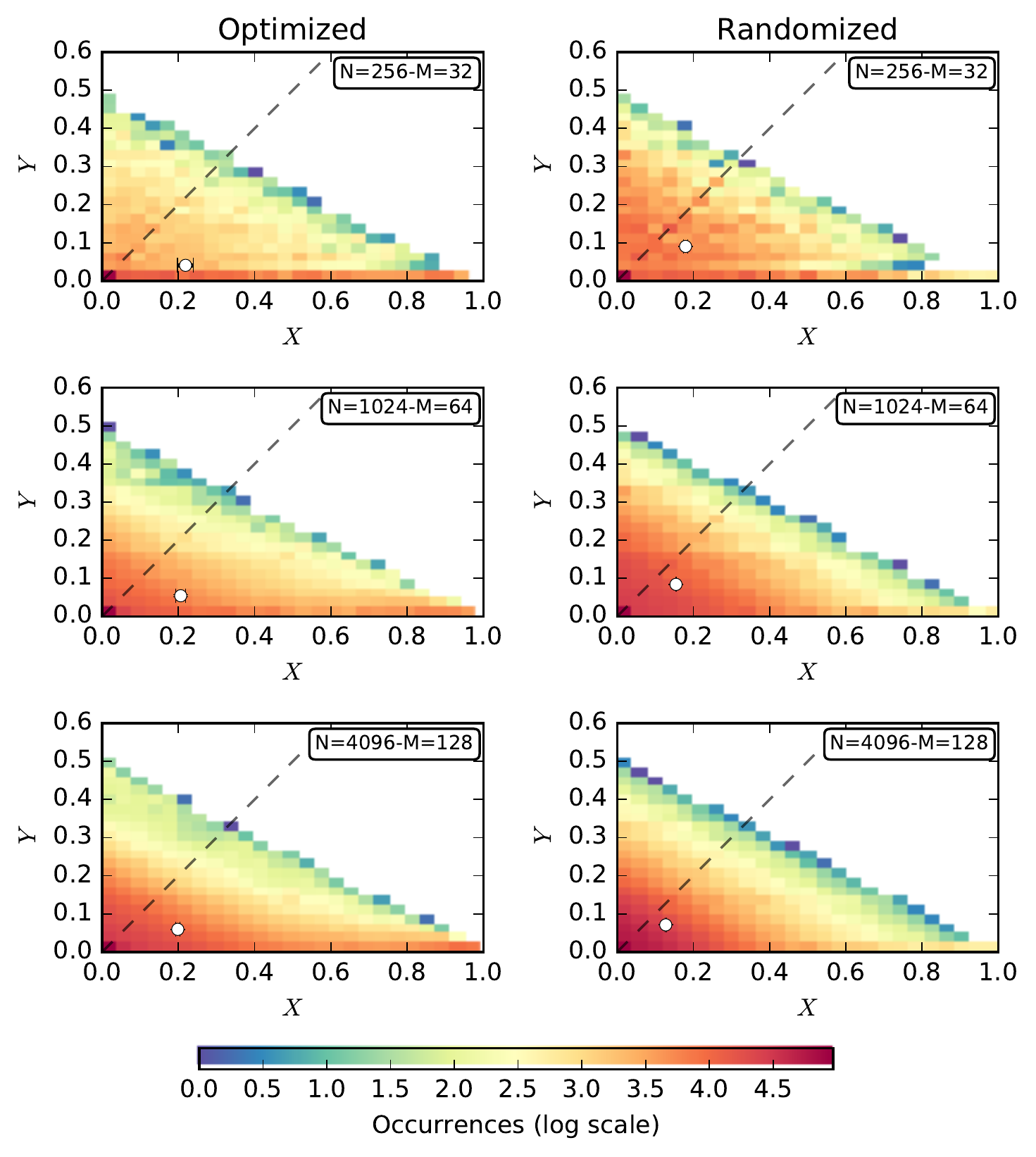}
\caption{Corner-constrained polymers: Every triple of distances (among homologue SAWs) contributes to the distribution $P(X,Y)$. We show $P(X,Y)$ for the optimized system (left) and for randomized paths (right): The standard error of the CM (over $10$ quenched disorders) compares with symbol size. A similar scenario (not shown) holds for the unconstrained polymers.}
\label{fig:triangles_tree}
\end{figure}

To further investigate this issue, we study the distribution of distance triples for the fixed-end polymer system. In Fig.~\ref{fig:triangles_tree} we plot the $P(X,Y)$ (considering all quenched disorders) both for optimized and randomized polymers with increasing lattice size. Plots for the single quenched disorders are shown in Fig. S8.
All optimized systems reveal higher concentrations (dark red) of triangles along the $X$-axis that is the signature of true UM; for larger $N$, the high $Y$ region gets progressively depleted (green-yellow).
The randomized plots show a $P(X,Y)$ that decreases with a constant gradient starting from a maximum in the origin (equilaterals). The $P(X,Y)$ mode is in all six cases in $(0,0)$, but for the optimized systems its value is roughly half the sum of bins along the $X$-axis.
This is better shown by the white circles representing the center of mass (CM) for each distribution: with respect to the randomized states, in which the CM shifts toward the origin with $N\rightarrow\infty$, the optimized states tend, on average and for each quenched disorder, to stay just above the $X$-axis without converging to $(0,0)$. The fraction of trivial UM is due to \textquotedblleft pathological\textquotedblright SAWs with very few ground states. Scalene triangles are produced by the shortest-path degeneracy due to local low-polymer density on the lattice (see the individual overlap distributions within the Supplemental Materials for details).


\paragraph*{Conclusions  --}
In this letter we presented a novel approach to explore the ground states of an important class of transport optimization problems on a regular square lattice. We showed that with this method one is able to obtain solutions for the optimal transport of a set of interacting communications spread over the lattice with different topological constraints. The interaction among paths is obtained within the unifying framework of concave/convex cost functions. The fact that this method works on a lattice, allows the possibility of discovering a hierarchy of the most inexpensive network infrastructures encouraging transport coalescence ($\gamma<1$) and then to optimize the distilled graph for performance, fault tolerance and congestion resistance with a repulsive cost function ($\gamma>1$) as in Ref. \cite{Yeung2013a}.
We tested our optimization procedure by characterizing the global cost gain and path-length variation for a system of randomly spread point-to-point communications over a square lattice that has been recently studied on random graphs via an unconventional use of the replica and cavity methods \cite{Yeung2013a} obtaining qualitatively similar results. 
The appearance, for $\gamma<1$, of families of hierarchically related solutions (tree-like as predicted by Banavar et al.\cite{Banavar2000}) led us to investigate whether RSB and an UM structure hold. Similarities and differences with respect to spin systems allowed us to borrow a standard approach to probe for UM, which we slightly extended by plotting the distribution of triangle types for growing lattice sizes.
In conclusion we found evidence supporting the RSB scenario (non-trivial overlap distributions) and UM at the level of single interacting polymers as hypothesized for DPRM systems~\cite{Zhang1987}. Here the equivalent of noise (highly correlated) is apparently played by all polymers minimizing the total energy thus forming a rough landscape~\cite{Buldyrev2006,Braunstein2002}. It should be further investigated whether and how this phenomenon depends on polymer density/dispersity and $\gamma$ values. Finally, we plan to further explore the possibility of defining a global similarity measure, encompassing all SAWs, to assess whether a system-wise UM structure exists.


\begin{acknowledgments}
We thank Francesco Versaci for useful discussions. We also acknowledge the contribution of Sardinian Regional Authorities under projects ABLE and Predict.
\end{acknowledgments}

\bibliography{PRE-bib,mypaperNotes}

\end{document}